\documentstyle[preprint,aps]{revtex}
\begin{document}
\draft
\title{
Note on the magnetotransport 
in the normal state of high-$T_{\rm c}$ cuprates
}
\author{
O. Narikiyo
}
\address{
Department of Physics, 
Kyushu University, 
Fukuoka 810-8560, 
Japan
}
\date{
Received \ \ \ \ \ \ \ \ \ \ \ \ \ 2000
}
\maketitle
\begin{abstract}
Theories of the magnetotransport, 
based on the quasipartricle of the Fermi-liquid, 
in the normal state of high-$T_{\rm c}$ cuprate superconductors, 
are critically examined and 
the necessity of the collective transport theory 
beyond the quasiparticle is discussed. 

\noindent
\end{abstract}
\newpage
\narrowtext
\sloppy
\maketitle
\section{Introduction}

  Anomalous temperature dependence 
of the magnetotransport coefficients 
in the normal state of high-$T_{\rm c}$ cuprate superconductors 
has been discussed intensively 
as the evidence for the breakdown of the Fermi-liquid  theory.\cite{Anderson1,Anderson2,Anderson3,Coleman1,Coleman2,Coleman3,Coleman4}    
  Experimental data are roughly summarized as\cite{Ex1,Ex2} 
$\sigma_{xx} \propto T^{-1}$,
$\sigma_{xy} \propto H T^{-3}$ and
$\Delta\sigma_{xx} \propto H^2 T^{-5}$
where $T$ is the temperature and
$H$ is the external magnetic field. 
  Thus Kohler's rule, expected from the ordinary transport theory 
based on the quasiparticle of the Fermi liquid, 
is strongly violated in a {\it wide} temperature range. 

  On the other hand, it has been claimed\cite{SP1,SP2,SP3,YaYa,KKU} 
that such a violation of Kohler's rule 
can be derived from the quasiparticle transport 
within the framework of the Fermi-liquid theory 
if the momentum-dependence of the scattering due to 
antiferromagnetic spin fluctuation is fully taken into account.  

  In this paper we show that the quasiparticle-transport theory 
is inadequate to explain the anomaly in section II 
and that the collective-transport theory beyond the quasiparticle 
is necessary in section III. 

\section{Quasiparticle Transport}

  The quasiparticle contribution\cite{Ex1,Ong,KSV} 
to the magnetotransport coefficients 
within the relaxation-time approximation 
in two dimensions 
is expressed in terms of the mean free path $l(s)$:   
\begin{eqnarray}
 \sigma_{xx} &=& e^2 
                 \int{\rm d}s\  l(s) [\cos\theta(s)]^2, \\
 \sigma_{xy} &=& e^2 \omega_{\rm c}
                 \int{\rm d}s\  l(s)\cos\theta(s)
                  {{\rm d}\over{\rm d}s}[l(s)\sin\theta(s)], \\
 \Delta\sigma_{xx} &=& - e^2 \omega_{\rm c}^2
                       \int{\rm d}s\  l(s)
                        \left\{ {{\rm d}\over{\rm d}s}
                                 [l(s)\cos\theta(s)] \right\}^2, 
\end{eqnarray}
where ${\rm d}s$ is the line element along the Fermi surface, 
$\theta(s)$ is the angle specifying the position on the Fermi surface 
and $\omega_{\rm c}$ is the cyclotron frequency. 
  Here the mean free path is proportional 
to the transport life-time $\tau_{\rm tr}(s)$. 
  The effect of the vertex correction 
beyond the relaxation-time approximation 
is discussed in Appendix A. 

  In order to obtain analytic results 
we use a model for the mean free path 
\begin{equation}
l(s)=l(\theta)=l_{\rm hot}{1+a \over 1+a\cos 4\theta},
\end{equation}
employed in the previous studies\cite{SP2,SP3}  
  Here $l_{\rm hot}\equiv l(\theta=0)$ and $a=(1-r)/(1+r)$ 
with $r=l_{\rm hot}/l_{\rm cold}$ and $l_{\rm cold}\equiv l(\theta=\pi/4)$. 
  From this model we can know 
qualitative features of the quasiparticle transport,  
though it is insufficient for quantitative discussions. 

  For this model replacing ${\rm d}s$ by ${\rm d}\theta$ we obtain 
\begin{eqnarray}
 \sigma_{xx} &=& 16\pi e^2 l_{\rm hot} r^{-1/2}, \\
 \sigma_{xy} &=& 8\pi e^2 \omega_{\rm c} l_{\rm hot}^2 (r+1)r^{-3/2}, \\
 \Delta\sigma_{xx} &=& - {\pi \over 8} e^2 \omega_{\rm c}^2
                       l_{\rm hot}^3 (-5r^4+52r^3+34r^2+52r-5)r^{-7/2}.  
\end{eqnarray}
  While eqs.~(2.5) and (2.6) agree with the previous result,\cite{SP2} 
eq.~(2.7) corrects the previous result\cite{SP3} 
as discussed in Appendix B.  

  If $r$ does not depend on the temperature, 
eqs.(2.5)-(2.7) satisfy Kohler's rule. 
  Although some violation of the rule can be derived 
from the temperature-dependence of $r$, 
it is impossible to explain the experimentally observed violation 
in a {\it wide} temperature range. 
  Namely it is impossible to obtain 
$(r+1)r^{-1/2} \propto T^{-1}$ 
and $(-5r^4+52r^3+34r^2+52r-5)r^{-2} \propto T^{-2}$, 
from the scattering due to antiferromagnetic spin fluctuation, 
in a {\it wide} temperature range consistent with the experiments. 
  Such a criticism has also been made by other authors\cite{OA} 
in another context. 

  Another criticism should be made for the case 
where the mean free path $l(s)$ is calculated 
from the phenomenological spin susceptibility.\cite{SP1,SP2,SP3,YaYa} 
  In those studies the self-consistency 
between the susceptibility and the self-energy for electrons 
has been neglected. 
  The requirement of the self-consistency 
weaken the temperature dependence of the Hall constant 
expected from the phenomenology\cite{SP1,SP2,SP3,YaYa} 
as has been shown by the numerical study 
based on the fluctuation-exchange approximation.\cite{KKU} 
  Consequently the temperature dependence of the Hall constant 
within the relaxation-time approximation 
is far weaker than that observed by the experiments.  

  In conclusion, 
only weak violation of Kohler's rule 
can be derived from the momentum-dependence of the transport life-time 
within the quasiparticle transport described by the Boltzmann equation. 

\section{Collective Transport}

  By the discussions in \S 2 
it has become clear that the anomalous temperature dependence 
observed in the normal state of high-$T_{\rm c}$ cuprate superconductors 
cannot be explained within the quasiparticle transport. 
  However, it does not directly lead to the breakdown\cite{Anderson1,Anderson2,Anderson3,Coleman1,Coleman2,Coleman3,Coleman4} 
of the Fermi-liquid theory. 
  There exists the collective contribution beyond the quasiparticle 
within the linear response theory based on the Fermi liquid: 
for example, 
the contributions C and D to the Hall conductivity 
discussed in ref.~18 correspond to the collective ones, 
while the contributions A and B 
constitute the quasiparticle transport theory 
consistent with the Boltzmann equation. 
  The collective contributions arise in the presence of the magnetic field 
so that Kohler's rule is strongly violated 
in a {\it wide} temperature range.\cite{Ando} 
  
  The collective transport theory is well established 
in the case of superconducting fluctuation.\cite{FET,BMT} 
  In the case of antiferromagnetic spin fluctuation 
relevant to the cuprate superconductors, 
the collective transport theory has been developed 
by the present authors\cite{MN1,MN2,MN3,MN4,NM} 
in parallel with the case of superconducting fluctuation. 
(See Appendix C.)

  The collective transport is necessary in the case of cuprates, 
since the anomalous phase with broken Kohler's rule 
is proximity to the antiferromagnetic insulator 
where the low-energy excitation is the spin wave 
and there is no quasiparticle at low energies. 
  It should be noted that the collective contributions are minor ones 
unless the spin fluctuation has the nesting character.\cite{MN1,MN2,MN3,MN4,NM} 
  The heart of the collective transport for the nested spin fluctuation 
is that the anomaly is regarded as $4k_{\rm F}$ singularity 
of the charge fluctuation triggered by two modes 
of the spin fluctuation with $2k_{\rm F}$ singularity.\cite{MN1} 
  The $4k_{\rm F}$ singularity is observed 
by the diffusive X-ray scattering experiment.\cite{Xray} 

  In conclusion, 
the strong violation of Kohler's rule 
should be ascribed to the collective contribution 
beyond the quasiparticle contribution described by the Boltzmann equation.  
 
\section{Summary}

  We have shown that the anomalous temperature dependence of 
the magnetotransport coefficients 
in the normal state of high-$T_{\rm c}$ cuprate superconductors 
is derived, 
within the framework of the Fermi-liquid theory, 
from the collective contribution of the nested spin fluctuation 
beyond the quasiparticle contribution 
described by the Boltzmann equation. 

  In addition, 
other transport coefficients should have the collective contribution 
from the nested spin fluctuation. 
  For example, the thermo-electric power 
has such a contribution\cite{MN2} 
in parallel with the case of the superconducting fluctuation.\cite{VL} 

\section*{Acknowledgements}

  The author has benefitted much from discussions with 
K. Miyake, C. M. Varma, K. Yamada, H. Fukuyama and H. Kohno 
in the last decade. 

\appendix
\section{Vertex Correction}

  In this Appendix we discuss the effect of the vertex correction. 
  In the case, where the Fermi-liquid theory is applicable, 
the current vertex, the three-point vertex appearing in Fig.~1, 
is determined by three contributions, Figs.~1(a)-1(c).\cite{YY} 
  This vertex correction leads to the collision term $I(n_k)$ 
in the Boltzmann equation\cite{Eliashberg,AGD,Okabe} 
with obvious notations: 
\begin{eqnarray}
 I(n_k) = &-&\sum_{k',q}W(k,k';k-q,k'+q)
             \delta(\varepsilon_k+\varepsilon_{k'}
                   -\varepsilon_{k-q}-\varepsilon_{k'+q}) \nonumber \\
          &\times&[n_k n_{k'}(1-n_{k-q})(1-n_{k'+q})
                  -(1-n_k)(1-n_{k'})n_{k-q} n_{k'+q}].
\end{eqnarray}
  This collision term is rewritten as\cite{Taylor,Chambers,Ziman,AM}
\begin{equation}
 I(n_k) = \sum_{k'} Q(k,k')[n_{k'} - n_k],
\end{equation}
with obvious notations. 
  The relation of the detailed balance, $Q(k,k')=Q(k',k)$, 
is satisfied if the collision term has the form 
given by eq.~(A.1).  

  In the absence of the magnetic field 
the Boltzmann equation leads to 
\begin{equation}
 {\vec v}_k = \sum_{k'} Q(k,k')[{\vec \Lambda}_k - {\vec \Lambda}_{k'}],
\end{equation}
where ${\vec v}_k$ is the velocity 
and ${\vec \Lambda}_k$ has the meaning of the mean free path. 
  Equation (A.3) is equivalent to eq.~(6.17) in ref.~29. 
  It is widely recognized\cite{Taylor,Chambers,Ziman,AM} that 
${\vec \Lambda}_k$ is not proportional to ${\vec v}_k$ in general, 
while it is stressed in ref.~14. 

  In the relaxation-time approximation,\cite{Ziman,AM} 
the right hand side of eq.~(A.2) is replaced by $-[n_k - n_k^0]/\tau_{\rm tr}(k)$ 
introducing the transport life-time $\tau_{\rm tr}(k)$ 
where $n_k^0$ is the equilibrium value.  
  In this approximation 
the effect of the vertex correction is partly taken into account. 
  However, it is shown\cite{YY,MF} that the full account of the vertex correction 
does not alter the temperature dependence of the resistivity 
obtained in the relaxation-time approximation.   

  In the presence of the magnetic field, 
the linearized Boltzmann equation for $g_k \equiv n_k - n_k^0$ is given by 
\begin{equation}
 e{\vec E}\cdot{\vec v}_k {\partial n_k^0 \over \partial \varepsilon_k}
 +{e \over c}({\vec v}_k \times {\vec B})\cdot{\partial g_k \over \partial{\vec k}}
 = \sum_{k'}Q(k,k')[g_{k'}-g_k].
\end{equation}
  This equation is rewritten in the matrix form:  
\begin{equation}
 \sum_{k'}A_{kk'}g_{k'}=
 -e{\vec E}\cdot{\vec v}_k {\partial n_k^0 \over \partial \varepsilon_k},
\end{equation}
where 
\begin{equation}
 A_{kk'} \equiv
 (\tau_{\rm tr}^{-1})_{kk'}
 -{e \over c}({\vec v}_k \times {\partial \over \partial{\vec k}})\cdot{\vec B}
  \delta_{k,k'},
\end{equation}
with 
\begin{equation}
 (\tau_{\rm tr}^{-1})_{kk'} \equiv
  {1 \over \tau_k}\delta_{k,k'} - Q(k,k'),
\end{equation}
and 
\begin{equation}
 {1 \over \tau_k}\equiv \sum_{k'} Q(k,k').
\end{equation}
  Since $g_{k'}$ is obtained by the matrix inversion: 
\begin{equation}
 g_{k'}=-\sum_{k}(A^{-1})_{k'k}
 e{\vec E}\cdot{\vec v}_k {\partial n_k^0 \over \partial \varepsilon_k},
\end{equation}
and the current ${\vec j}$ is obtained by 
\begin{equation}
 {\vec j}=e\sum_{k'}{\vec v}_{k'}g_{k'},
\end{equation}
the transport coefficient $\sigma_{\mu\nu}$ is given as\cite{KSV,Taylor} 
\begin{equation}
 \sigma_{\mu\nu}= - e^2\sum_{kk'}
 v_k^\mu (A^{-1})_{k'k} v_{k'}^\nu {\partial n_k^0 \over \partial \varepsilon_k},
\end{equation}
and this result fully contains the effect of the vertex correction. 
  Our previous result\cite{NM} is consistent with eq.~(A.11) but 
the momentum-derivative of the transport life-time has been omitted 
in eq.~(3.38) of ref.~18 and thus in eq.~(21) of ref.~14. 

  Using the fluctuation-exchange (FLEX) approximation, 
the authors of ref.~14 claim that 
the vertex correction in the presence of strong antiferromagnetic 
spin fluctuation leads to strong temperature dependence 
of the transport coefficients 
unexpected from the relaxation-time approximation. 
  While their study is in the framework of the Fermi-liquid theory, 
such a claim contradicts with the previous studies\cite{KY,YY} 
based on the Fermi-liquid theory. 
  In the following 
we show that the FLEX approximation does not give 
the correct vertex correction discussed above. 

  In order to obtain the correct vertex correction, eq.~(A.1), 
consistent with the Pauli exclusion principle for fermions,   
at least following two conditions should be satisfied. 
  
  First, the four-point vertex appearing in Fig.~1 
should satisfy the so-called crossing or exchange symmetry. 
  In the previous studies\cite{KY,YY} based on the Fermi-liquid theory 
this symmetry is assumed. 
  For example, the four-point vertex $\Gamma_{\rm ph}$ 
in the particle-hole channel shown in Fig.~2(a) and 
$\Gamma_{\rm pp}$ in the particle-particle channel shown in Fig.~2(b) 
are assumed to be identical. 
  On the other hand, in the FLEX approximation 
such a symmetry is broken.\cite{BW,DHS,VT} 
  For example, the ladder process typical in the FLEX approximation 
shown in Fig.~3 cannot be identical in the sense discussed above. 

  Second, only low-energy states 
described by the singular part of the electron Green function 
should appear 
connecting each vertices 
in the vertex correction of the Fermi-liquid shown in Fig.~1. 
  This limitation ensures that the resistivity is proportional to $T^2$ 
in the ordinary Fermi-liquid theory. 
  On the other hand, in the FLEX approximation\cite{KKU} 
this limitation is violated in the so-called Maki-Thompson process 
which has no counterpart in the Fermi-liquid theory. 
  Namely, the Maki-Thompson process 
contains unphysical high-energy states 
in comparison with the process of Fig.~1(a). 
  While the Maki-Thompson process is treated on equal footing 
with the so-called Aslamazov-Larkin processes 
corresponding to Figs.~1(b) and 1(c) 
in the Fermi-liquid theory 
and such a treatment guarantees the vanishing resistivity 
in the absence of Umklapp scattering 
as stressed in ref.~29, 
the Maki-Thompson process is overestimated in the FLEX approximation. 
  It is natural to expect that the fact, 
that the Aslamazov-Larkin processes 
have little effect on the transport coefficients, 
found in the FLEX approximation\cite{KKU} 
supports the validity of the relaxation-time approximation. 

  The reason, why the vertex correction in Fig.~1 plays little role 
besides the introduction of the transport life-time, 
is obvious from the previous study:\cite{YY} 
the contribution from the two four-point vertices and two Green functions 
connecting them, and the one from the two Green functions 
connecting the four-point vertex and the three-point vertex, 
both of which are related to the imaginary part of the electron selfenergy, 
cancel out and the remaining effect is mostly taken into account 
by introducing the transport life-time. 

  The theoretical results by the FLEX approximation 
cannot be compatible with the experiments at least by two reasons. 
  First, the magnetotransport anomaly 
in the normal state of high-$T_{\rm c}$ cuprate superconductors 
should be understood in a single framework 
which can explain the so-called pseudogap phenomena,\cite{Ex1,Ex2} 
while the FLEX approximation at present 
cannot explain the pseudogap. 
  Second, anomalous temperature dependences 
observed in several experiments\cite{Ando,RHCH} 
cannot be ascribed to the vertex correction. 

\section{Magnetoconductivity}

  The magnetoconductivity in the model used in \S 2 
is given by 
\begin{equation}
\Delta\sigma_{xx}=-16 e^2 \omega_{\rm c}^2 l_{\rm hot}^3 (1+a)^3
                  \int_0^\pi{\rm d}\theta
                  \left\{ {1 \over (1+a\cos\theta)^3}  
                         +{a^2\sin^2\theta \over (1+a\cos\theta)^5} \right\}.
\end{equation}
  These two integrals can be evaluated 
by use of the parameter-derivative of the formulae\cite{GR} 
\begin{eqnarray}
 \int_0^\pi {\rm d}x {1 \over p+q\cos x}
  &=& \pi (p^2-q^2)^{-1/2}, \\
 \int_0^\pi {\rm d}x {\sin^2 x \over p+q\cos x}
  &=& \pi [p-(p^2-q^2)^{1/2}]q^{-2},  
\end{eqnarray}
for $p > |q|$ 
and the result is given in \S 2. 
  In the previous study\cite{SP3} the second integral in eq.~(B.1) 
has been omitted. 

\section{Sign of Hall Conductivity}

  The collective contribution of the Aslamazov-Larkin process 
of the superconducting fluctuation to the Hall conductivity 
is proportional to $N_{\rm F}'$ 
where $N_{\rm F}'$ is the energy derivative 
of the density of states at the Fermi energy. 
(The definition of $\alpha$ in the right hand side of eq.~(2.27) 
in ref.~20 should be multiplied by minus sign.) 
  On the other hand, 
the one of the nested spin fluctuation 
is proportional to $-N_{\rm F}'$. 
  The difference of the sign can be understood as follows. 
  In the case of the superconducting fluctuation 
the diagram for the Hall conductivity of the Aslamazov-Larkin process 
is derived from Fig.~4(a) by attaching three current vertices to it. 
  In this process 
the electron with the momentum $p$ 
and the fermionic frequency $\varepsilon_n$ 
interacts with the hole with $-p$ and $-\varepsilon_n$ 
and feels a reduced magnetic field by the motion of the hole. 
  If we consider free electrons,  
$N_{\rm F}'$ is positive and so that the Hall conductivity 
due to the superconductive fluctuation is positive, 
while the Hall conductivity due to the quasiparticle is negative. 
  In the case of nested spin fluctuation 
the Aslamazov-Larkin process is derived from Fig.~4(b) 
where the electron with $p$ and $\varepsilon_n$ 
interacts with the electron with $p+Q$ and $\varepsilon_n$. 
  Since $p+Q \sim -p$ with the nesting vector $Q$, 
each electron moving in the opposite direction 
feels an induced magnetic field. 
  It should be noted that we need the vertex correction 
in order to obtain non-vanishing collective contribution 
in the case of the spin fluctuation.\cite{MN3} 



\begin{figure}
\caption{
The current-vertex corrections. 
The open square represents the interaction vertex and 
the open triangle the current vertex. 
}
\label{fig:1}
\end{figure}

\begin{figure}
\caption{
The interaction vertex (a) in the particle-hole channel 
and (b) in the particle-particle channel. 
}
\label{fig:2}
\end{figure}

\begin{figure}
\caption{
The second order vertex (a) in the particle-hole channel 
and (b) in the particle-particle channel. 
}
\label{fig:3}
\end{figure}

\begin{figure}
\caption{
The generating function of the magnetoconductivity 
(a) for the superconducting fluctuation and 
(b) for the spin fluctuation. 
}
\label{fig:4}
\end{figure}


\end{document}